\documentstyle[12pt,twoside,fleqn,qm96]{article}

\newcommand{\AmS}{{\protect\the\textfont2
  A\kern-.1667em\lower.5ex\hbox{M}\kern-.125emS}}

\title{How to extract physics from HBT radius parameters}

\author{Ulrich Heinz\address{Physics Department, Duke University,
                             Durham, NC 27708-0305, USA}%
       \thanks{On sabbatical leave from Institut f\"ur 
               Theoretische Physik, Universit\"at Regensburg, D-93040 
               Regensburg, Germany. -- Invited talk given at Quark Matter 
               '96, Heidelberg, 20.-24.5.96} 
       }

\begin{document}
% typeset front matter
\maketitle

\begin{abstract}
 {\small \noindent
 I review recent progress in the understanding of the connection 
 between the space-time structure of the particle emitting source
 and the form of the two-particle correlation function in momentum
 space. Based on a new scheme for calculating the HBT radius 
 parameters from the emission function, strategies are suggested to 
 separate for rapidly expanding sources the information on the spatial 
 and temporal structure of the source. To this end a new fitting 
 function for the two-particle correlation function is proposed. Its 
 usefulness is demonstrated for a typical expanding model source, and 
 it is shown how the dependence of the resulting fit parameters on the 
 momentum of the particle pair can be used to measure the longitudinal 
 and transverse expansion of the source.  
 }
\end{abstract}

%%%%%%%%%%%%%%%%%%%%%%%%%%%%%%%%%%%%%%%%%%%%%%%%%%%%%%%%%%%%%%
\section{INTRODUCTION}
\label{sec0}
%%%%%%%%%%%%%%%%%%%%%%%%%%%%%%%%%%%%%%%%%%%%%%%%%%%%%%%%%%%%%%

The only known way to obtain direct experimental information on the 
space-time structure of the particle emitting source created in a
relativistic nuclear collision is through two-particle intensity
(Hanbury-Brown--Twiss (HBT)) interferometry \cite{BGJ90}. The 
goal of this method is to extract the {\em space-time} structure of 
the source from {\em momentum spectra} which are the only measurable 
quantities, making use of the quantum statistical correlations between 
pairs of identical particles. This information is crucial for an 
assessment of theoretical models which try to extract the energy 
density of the source from the measured single particle spectra and 
particle multiplicity densities in momentum space. Reliable estimates 
of the source volume and the energy density are, on the other hand, 
indispensable for an experimental proof that high energy collisions can 
successfully generate large volumes of matter with extreme energy 
density, where a transition into deconfined quark matter might be 
possible.  

For many years HBT interferometry of hadron-hadron and nucleus-nucleus
collisions was motivated by the naive expectation, based on the 
experience with photon interferometry for stars, that the width of the 
two-particle correlation function in the relative momentum directly 
measures the geometric size of the source. This expectation is {\em 
wrong}. Unlike stars in the universe, the sources created in hadronic 
or heavy-ion collisions may feature inhomogeneous temperature profiles 
and strong collective dynamical expansion. We now know 
\cite{MS88,MSH92,AS95,CL95,CSH95} that for such sources the HBT 
radius parameters generally don't measure the full source size, but 
only so-called ``space-time regions of homogeneity" \cite{MS88,AS95} 
inside which the momentum distribution varies sufficiently little so 
that the particles can actually show the quantum statistical 
correlations. The size of these homogeneity regions varies with the 
momentum of the emitted particles, causing a dependence of the HBT
parameters on the pair momentum 
\cite{MS88,MSH92,AS95,CL95,CSH95,P84,B89,PCZ90,Marb,CNH95,WSH96,HTWW96,WHTW96}. 
The detailed momentum dependence is, however, model-dependent, and in 
general it is not simple \cite{WSH96}. The extraction of the strength 
of the collective flow from HBT data is further complicated by 
significant resonance decay contributions which also induce a 
momentum dependence of the HBT radius parameters and of the so-called 
``incoherence parameter" \cite{Marb,CLZ94}.  

The finite lifetime of the sources created in nuclear collisions leads 
to another complication: the HBT radius parameters generally mix the 
spatial and temporal aspects of the source extension in a non-trivial 
and reference frame dependent way, in particular if at the same time 
the source undergoes collective expansion. The origin and pattern of 
this mixing was clarified in a recent series of publications from the 
Regensburg group \cite{CSH95,CNH95,WSH96,HTWW96,WHTW96}. Two major new 
results have resulted from this work: (i) the discovery \cite{CSH95} 
of a new cross term in the two-particle correlation function mixing 
the outward and longitudinal components of the relative momentum 
between the two particles, and (ii) the new YKP 
(Yano-Koonin-Podgoretski\u\i) parametrisation of the correlation 
function \cite{CNH95,HTWW96,WHTW96}. The latter permits the 
experimental determination of the longitudinal velocity of the source 
volume element where most of the particle pairs originate (as a 
function of the pair momentum) and achieves a nearly perfect 
factorization of the longitudinal, transverse and temporal homogeneity 
regions of the source (again as functions of the pair momentum) in the 
source rest frame. Furthermore it provides for a clean separation of the 
longitudinal and transverse dynamics of the source.  

In this talk I review these new theoretical developments and 
exemplify them for a class of simple model emission functions for 
thermalized sources with collective transverse and longitudinal 
expansion and finite space-time geometry.

%%%%%%%%%%%%%%%%%%%%%%%%%%%%%%%%%%%%%%%%%%%%%%%%%%%%%%%%%%%%%%
\section{EMISSION FUNCTION AND PARTICLE SPECTRA}
\label{sec1}
%%%%%%%%%%%%%%%%%%%%%%%%%%%%%%%%%%%%%%%%%%%%%%%%%%%%%%%%%%%%%%

The single and two particle spectra are defined as
 \begin{eqnarray}
 \label{spectra1}
    P_1({\bf p}) &=& E_p {dN \over d^3p} 
                  = E_p \langle \hat a^+_p \hat a_p \rangle\, ,
 \\
 \label{spectra2}
    P_2({\bf p_1},{\bf p_2}) &=& E_1 E_2 {dN \over d^3p_1 d^3p_2} 
                  = E_1 E_2 \langle 
                            \hat a^+_{p_1} \hat a^+_{p_2}
                            \hat a_{p_2} \hat a_{p_1} 
                            \rangle
 \end{eqnarray}
in terms of creation and destruction operators for on-shell particles 
with momenta ${\bf p}_i$, where $\langle \dots \rangle$ denotes an 
average over the source ensemble. $P_1$ and $P_2$ are normalized to 
the average number of particles $\langle N \rangle$ and pairs
$\langle N(N-1)\rangle$ per event, respectively. The two-particle 
correlation function is defined as
 \begin{equation}
 \label{correl}
  C({\bf p_1},{\bf p_2}) = 
  {\langle N\rangle^2 \over \langle N(N-1) \rangle}\,
  {P_2({\bf p_1},{\bf p_2}) \over P_1({\bf p_1}) P_1({\bf p_2})}\, .
 \end{equation}
For uncorrelated emission and in the absence of final state 
interactions \cite{BB96} one can prove \cite{GKW79,CH94} a generalized 
Wick theorem for the factorisation of the 2-particle spectrum 
(\ref{spectra2}) and obtains
 \begin{equation}
 \label{correl1}
  C({\bf q},{\bf K}) = 1 \pm 
    {\left\vert \langle \hat a^+_{p_1} \hat a_{p_2} \rangle 
     \right\vert^2
     \over \langle \hat a^+_{p_1} \hat a_{p_1} \rangle
           \langle \hat a^+_{p_2} \hat a_{p_2} \rangle}
 \end{equation}
where ${\bf q} = {\bf p_1} - {\bf p_2}$ and 
${\bf K} = ({\bf p_1} + {\bf p_2})/2$ denote the relative and
total momentum of the particle pair, and the positive (negative) sign 
applies for bosons (fermions). Note that the second term is positive
definite.

These expressions can be further simplified and turned into a 
practical starting point for computations by introducing the emission 
function $S(x,K)$. It is defined in terms of the classical source 
amplitude $J(x)$ for creating a free pion state \cite{GKW79} via the 
Wigner transform of its associated density matrix 
 \begin{equation}
 \label{wigner}
   S(x,K) = \int {d^4y \over 2(2\pi)^3} e^{-K{\cdot}y}\, 
            \left\langle J^*\left(x+{\textstyle{y\over 2}}\right) 
                         J\left(x-{\textstyle{y\over 2}}\right) 
            \right\rangle
 \end{equation}
and is the quantum mechanical analogue of the classical phase-space 
density which gives the probability for creating a free particle with 
four-momentum $K$ at space-time point $x$. In terms of this emission 
function the single particle spectrum is given by
 \begin{equation}
 \label{single}
   E_K {dN\over d^3K} = \int d^4x\, S(x,K)
 \end{equation}
where the r.h.s. is to be evaluated on-shell, i.e. at $K^0 = E_K =
\sqrt{m^2 + {\bf K}^2}$. The two-particle correlation function
is obtained from \cite{S73,GKW79,P84,CH94}
 \begin{equation}
 \label{double}
  C({\bf q},{\bf K}) \approx 1 \pm 
    {\left\vert \int d^4x\, S(x,K)\, e^{iq{\cdot}x}
     \right\vert^2
     \over 
     \left\vert \int d^4x\, S(x,K) \right\vert^2} 
   = 1 \pm \left\vert\left\langle e^{iq{\cdot}x} \right\rangle
           \right\vert^2
 \end{equation}
where the r.h.s. must be evaluated at $q=p_1-p_2$, $K=(p_1+p_2)/2$
with $p_i$ on-shell. (This implies $K{\cdot}q=0$.) The approximation 
consists of replacing the single particle spectra at $p_1$ and $p_2$ 
in the denominator by the spectrum at $K=(p_1+p_2)/2$; it is exact for 
exponential momentum spectra and a good approximation in practice 
\cite{CSH95}. The second equality in (\ref{double}) defines a 
($K$-dependent) average over the emission function of which we will 
make abundant use below. A useful feature is that in (\ref{double})
the emission function can be to very good approximation 
\cite{CSH95,PCZ90} evaluated at $K^0=E_K$, i.e. on the classical 
energy shell, since the typical source radii are larger than the 
Compton wavelengths of the observed hadrons. This warrants the 
replacement of the Wigner density $S(x,K)$ by a classical phase-space 
distribution in practical calculations. 

Due to the on-shell constraint $q{\cdot}K=0$ the four components of 
$q$ are not independent, but related by
 \begin{equation}
 \label{massshell}
   q^0 = \bbox{\beta}\cdot {\bf q} \qquad {\rm with} \qquad 
   \bbox{\beta} = {{\bf K}\over K^0} \approx {{\bf K}\over E_K}\, .
 \end{equation} 
The Fourier transform in (\ref{double}) is therefore not invertible, 
and the reconstruction of the space-time structure of the source from 
HBT measurements will thus always require additional model 
assumptions. Furthermore, inserting (\ref{massshell}) into (\ref{double}),
$iq{\cdot}x = i {\bf q}\cdot ({\bf x} - \bbox{\beta}t)$, we see that the 
correlator $C({\bf q},{\bf K})$ mixes the spatial and temporal information
in a non-trivial way which depends on the pair velocity $\bbox{\beta}$.
Only for time-independent sources things become simple: the correlator
then measures the Fourier transform of the spatial source distribution,
however only in the directions perpendicular to $\bbox{\beta}$ since 
the time integration leads to a $\delta$-function $\delta(
\bbox{\beta}{\cdot}{\bf q})$.

From Eq.~(\ref{double}) it is clear that, unless the emission function 
factorizes in $x$ and $K$, $S(x,K) = F(x)G(K)$ (in which case $G(K)$
cancels between numerator and denominator), the correlator is a function 
of {\em both} ${\bf q}$ and ${\bf K}$. If one parametrises it by a 
Gaussian in $q$ (see Sec.~\ref{sec3}) this results in ${\bf K}$-dependent
width parameters (``HBT radii''). In thermal sources $x-K$ correlations 
which spoil such a factorization can be induced by temperature gradients 
and/or collective expansion (with a 4-velocity $u^\mu(x)$): in both cases 
the momentum spectrum $\sim \exp[-p{\cdot}u(x)/T(x)]$ of the 
emitted particles depends on the emission point.

%%%%%%%%%%%%%%%%%%%%%%%%%%%%%%%%%%%%%%%%%%%%%%%%%%%%%%%%%%%%%%
\section{MODEL-INDEPENDENT EXPRESSIONS FOR THE HBT RADII}
\label{sec2}
%%%%%%%%%%%%%%%%%%%%%%%%%%%%%%%%%%%%%%%%%%%%%%%%%%%%%%%%%%%%%%

One of the crucial questions is, of course, to what extent a measured
${\bf K}$-dependence of the HBT radii allows for a quantitative 
reconstruction of the collective source dynamics. To answer it we must 
first learn more about the physical meaning of these ``radii''.
To this end it is useful to write the emission function in the following
form \cite{CNH95,WSH96,HTWW96}:
 \begin{equation}
 \label{7}
   S(x,K) = N({\bf K})\, S(\bar x({\bf K}),K)\, 
            \exp\left[ - {1\over 2} \tilde x^\mu({\bf K})\, 
            B_{\mu\nu}({\bf K})\,\tilde x^\nu({\bf K})\right]
   + \delta S(x,K) \, ,
 \end{equation} 
where (with expectation values defined as in (\ref{double}))
 \begin{equation}
 \label{8}
  \bar x_\mu({\bf K}) = \langle x_\mu \rangle , \ \
  \tilde x_\mu ({\bf K}) = x_\mu - \bar x_\mu({\bf K}) , \ \
  (B^{-1})_{\mu\nu}({\bf K}) 
  = \langle \tilde x_\mu \tilde x_\nu \rangle .
 \end{equation}
This construction ensures that the term $\delta S$ has vanishing zeroth, 
first and second order moments and thus contains only higher order 
information on sharp edges, wiggles, secondary peaks, etc. in the source. 
It was shown numerically \cite{WSH96} to have negligible influence on 
the half width of the correlation function and to contribute only weak, 
essentially unmeasurable structures in $C({\bf K},{\bf q})$ at large 
values of ${\bf q}$. Neglecting $\delta S$, the two-particle correlation 
function (\ref{double}) can be calculated analytically: 
 \begin{equation}
 \label{11}
  C({\bf K},{\bf q}) = 1 + \exp\left[ - q^\mu q^\nu 
            \langle \tilde x_\mu \tilde x_\nu \rangle ({\bf K}) 
                  \right] \, .
 \end{equation}
Please note that the point $\bar x^\mu({\bf K})$ of maximum emissivity 
at momentum ${\bf K}$ is unmeasurable \cite{HTWW96,WH96}. Only the 
${\bf K}$-dependent effective widths (``lengths of homogeneity'') 
$\langle \tilde x_\mu \tilde x_\nu \rangle ({\bf K})$ of the source 
of particles with momentum ${\bf K}$ are accessible by HBT interferometry. 

Actually, due to the on-shell constraint (\ref{massshell}), only 6 linear 
combinations of the 10 variances $\langle \tilde{x}_\mu \tilde{x}_\nu 
\rangle({\bf K})$ are measurable \cite{CNH95}; in the case of 
azimuthal symmetry of the source around the beam axis, this number 
reduces to 4 out of 7. Which linear combinations occur in practice 
depends on the way the correlation function is parametrised. The general 
form (\ref{11}) together with (\ref{massshell}) still provide some 
freedom as to which components of $q$ to keep as independent variables 
(see Sec.~\ref{sec3}). But whichever choice one makes, all the 
${\bf K}$-dependent parameters (``HBT radii'') in the resulting 
Gaussian function of $q$ can be easily calculated from the variances 
$\langle \tilde x^\mu \tilde x^\nu \rangle$, i.e. by simple quadrature 
formulae, for arbitrary emission functions $S(x,K)$. The relation
between the HBT parameters and the variances is {\em model-independent},
i.e. it does not depend on the form of the emission function $S(x,K)$.

%%%%%%%%%%%%%%%%%%%%%%%%%%%%%%%%%%%%%%%%%%%%%%%%%%%%%%%%%%%%%%
\section{STANDARD AND YKP FITS TO THE CORRELATION FUNCTION}
\label{sec3}
%%%%%%%%%%%%%%%%%%%%%%%%%%%%%%%%%%%%%%%%%%%%%%%%%%%%%%%%%%%%%%

For the following discussion we employ the conventional \cite{P84,B89}
Cartesian coordinate system with $z$ along the beam axis and ${\bf K}$ 
lying in the $x$-$z$-plane. The $z$-component of a 3-vector is labelled 
by $l$ (for {\em longitudinal}), the $x$-component by $o$ (for 
{\em outward}) and the $y$-component by $s$ (for {\em sideward}). 
Then $\beta_s=0$ such that $q^0 = \beta_\perp q_o + \beta_l q_l$,
with $\beta_\perp = \vert {\bf K}_\perp \vert / K^0$ being 
(approximately) the velocity of the particle pair transverse to the 
beam direction while $\beta_l$ is its longitudinal component.

The standard Cartesian parametrization \cite{CSH95} of the 
correlation function is obtained by using this condition
to eliminate $q^0$ from Eq.~(\ref{11}). One obtains 
 \begin{equation}
     C({\bf K},{\bf q})
    = 1 + \exp\left[ -\sum_{i,j=s,o,l} R_{ij}^2({\bf K})\, q_i\, q_j 
              \right]
 \label{13}
 \end{equation}
where the 6 HBT ``radius parameters'' $R_{ij}$ are given as \cite{CSH95,HB95}
 \begin{equation}   
   R_{ij}^2({\bf K}) = 
   \langle (\tilde{x}_i-{\beta}_i\tilde{t})
           (\tilde{x}_j-{\beta}_j\tilde{t})\rangle \, ,
   \quad i,j = s,o,l \, ,
 \label{14}
 \end{equation}
 through
through the space-time variances of the source. For an azimuthally symmetric 
sample of collision events, $C({\bf q}, {\bf K})$ is symmetric with respect 
to $q_s \to -q_s$ 
\cite{CNH95}. Then $R_{os}^2 = R_{sl}^2 = 0$ and
 \begin{eqnarray}   
    C({\bf K},{\bf q})
    &=& 1 + \exp\left[ - R_s^2({\bf K}) q_s^2 - R_o^2({\bf K}) q_o^2
                     - R_l^2({\bf K}) q_l^2 - 2 R_{ol}^2({\bf K}) q_o q_l
              \right] \, , 
    \quad \text{with}
 \label{15}\\
   R_s^2({\bf K}) &=& \langle \tilde{y}^2 \rangle \, ,
 \label{16a}\\
   R_o^2({\bf K}) &=& 
   \langle (\tilde{x} - \beta_\perp \tilde t)^2 \rangle \, ,
 \label{16b}\\
   R_l^2({\bf K}) &=& 
   \langle (\tilde{z} - \beta_l \tilde t)^2 \rangle \, ,
 \label{16c}\\
   R_{ol}^2({\bf K}) &=& 
   \langle (\tilde{x} - \beta_\perp \tilde t)
           (\tilde{z} - \beta_l \tilde t) \rangle \, .
 \label{16d} 
 \end{eqnarray}
The cross-term (\ref{16d}) was only recently discovered \cite{CSH95}.
Clearly these HBT radius parameters mix spatial and temporal 
information on the source in a non-trivial way. Their interpretation 
in various reference systems, in particular the meaning of the 
generally non-vanishing cross-term $R_{ol}^2$, was extensively 
discussed in Refs.~\cite{CSH95,CNH95,WSH96}, by analysing 
these expressions analytically for a large class of (azimuthally 
symmetric) model source functions and comparing with the numerically 
calculated correlation function (\ref{double}). An important observation 
resulting from these studies is that the difference 
 \begin{equation}
 \label{17}
   R_{\rm diff}^2 \equiv  R_o^2 - R_s^2 =
   \beta_\perp^2 \langle \tilde t^2 \rangle - 2 \beta_\perp \langle
   \tilde{x} \tilde t\rangle + (\langle \tilde x^2 \rangle -
   \langle \tilde y^2 \rangle)
 \end{equation}
is generally dominated by the first term on the r.h.s. \cite{WHTW96} and thus 
provides access to the lifetime $\Delta t = \sqrt{\langle t^2 \rangle 
- \langle t \rangle^2}$ of the source \cite{CP91} (more exactly: the 
duration of the particle emission process). In heavy-ion 
collisions, due to rapid expansion of the source, one would generally
not expect $\langle \tilde t^2 \rangle$ to be much larger than
either $\langle \tilde x^2 \rangle$ or $\langle \tilde y^2 \rangle$ 
(see however \cite{RG96} for possible exceptions near a phase transition to
QGP). In the standard fit one is not sensitive to small values of $\Delta t$
since Eq.~(\ref{17}) then involves a small difference of two large
numbers, each associated with standard experimental errors. The
factor $\beta_\perp^2 \leq 1$ in front of $\langle \tilde t^2 \rangle$
further complicates its extraction, in particular at low $K_\perp$ 
where $\Delta t({\bf K})$ is usually largest (see below). 

This problem is avoided in the Yano-Koonin-Podgoretski\u\i\ parametrisation
\cite{YK78,P83,CNH95,HTWW96,WHTW96} of the correlation function for 
azimuthally symmetric systems. It is based on an elimination 
of $q_o$ and $q_s$ in terms of $q_\perp^2 = q_o^2 + q_s^2$, 
$q^0$, and $q_3$ in (\ref{11}):
 \begin{equation}
 \label{18}
   C({\bf q},{\bf K}) =
       1 +  \exp\left[ - R_\perp^2\, q_{\perp}^2 
                       - R_\parallel^2 \left( q_l^2 - (q^0)^2 \right)
                       - \left( R_0^2 + R_\parallel^2 \right)
                         \left(q{\cdot}U\right)^2
                \right]  ,
 \end{equation}
with four ${\bf K}$-dependent parameters $R_\perp$, $R_\parallel$, 
$R_0$, and $U^\mu$ where the latter is a 4-velocity with only a 
longitudinal spatial component:
 \begin{equation}
 \label{19}
   U({\bf K}) = \gamma({\bf K}) \left(1, 0, 0, v({\bf K}) \right) ,
   \ \ \text{with} \ \
   \gamma = (1 - v^2)^{-1/2}\, .
 \end{equation}
This parametrisation has the advantage that the fitted YKP parameters 
$R_\perp^2({\bf K})$, $R_\parallel^2({\bf K})$, and $R_0^2({\bf K})$ 
do not depend on the longitudinal velocity of the observer system.
They (as well as $v({\bf K})$) can be calculated from the variances 
$\langle \tilde x^\mu \tilde x^\nu \rangle$ in any reference frame 
(see \cite{HTWW96} for explicit expressions), but their physical 
interpretation is easiest in terms of coordinates measured in the 
frame where $v({\bf K})$ vanishes. There they are given by 
\cite{CNH95} 
 \begin{eqnarray}   
   R_\perp^2({\bf K}) &=& R_s^2({\bf K}) = \langle \tilde{y}^2 \rangle \, ,
 \label{20a} \\
   R_\parallel^2({\bf K}) &=& 
   \left\langle \left( \tilde z - \beta_l \tilde x/\beta_\perp \right)^2 
   \right\rangle   
     - \beta_l^2 \langle \tilde y^2 \rangle / \beta_\perp^2 
     \approx \langle \tilde z^2 \rangle \, ,
 \label{20b} \\
   R_0^2({\bf K}) &=& 
   \left\langle \left( \tilde t - \tilde x/\beta_\perp \right)^2 
   \right\rangle 
    - \langle \tilde y^2 \rangle/\beta_\perp^2 
    \approx \langle \tilde t^2 \rangle \, ,
 \label{20c}
 \end{eqnarray}
where in the last two expressions the approximation consists of 
dropping generically small \cite{CNH95} terms (for a quantitative 
discussion see \cite{WHTW96}). The first expression (\ref{20a}) 
remains true in any longitudinally boosted frame.

Eq.~(\ref{20c}) shows that the YKP parameter $R_0({\bf K})$ 
measures directly (up to the neglected small terms) the time 
duration $\Delta t({\bf K})$ during which particles of momentum 
${\bf K}$ are emitted, in the frame were the YKP velocity $v({\bf K})=0$. 
The advantage compared to the standard Cartesian fit is 
that here it is fitted directly, and no problems of differences of large 
numbers occur in its extraction. 

Since the standard Cartesian and YKP parametrizations (\ref{15}) and 
(\ref{18}) of the correlator differ only by the choice of independent 
components of $q$, the two sets of HBT parameters must be related. One finds
\cite{HTWW96}
 \begin{eqnarray}
 \label{24z}
   R_s^2 &=& R_\perp^2\, ,
 \\
 \label{24a}
   R_{\rm diff}^2 &=& R_o^2 - R_s^2 = \beta_\perp^2 \gamma^2 
             \left( R_0^2 + v^2 R_\parallel^2 \right) \, ,
 \\
 \label{24b}
   R_l^2 &=& \left( 1 - \beta_l^2 \right) R_\parallel^2 
             + \gamma^2 \left( \beta_l-v \right)^2
             \left( R_0^2 + R_\parallel^2 \right)\, ,
 \\
 \label{24c}
   R_{ol}^2 &=& \beta_\perp \left( -\beta_l R_\parallel^2 
             + \gamma^2 \left( \beta_l-v \right)^2
             \left( R_0^2 + R_\parallel^2 \right) \right)\, .
 \end{eqnarray}
These relations provide a powerful consistency check on the experimental 
fitting procedure of the correlation function, of similar value as the 
relation \cite{CNH95,WSH96} $\lim_{K_\perp \to 0} (R_o({\bf K}) - 
R_s({\bf K})) = 0$ which results from azimuthal symmetry.  

%%%%%%%%%%%%%%%%%%%%%%%%%%%%%%%%%%%%%%%%%%%%%%%%%%%%%%%%%%%%%%
\section{A SIMPLE SOURCE MODEL}
\label{sec4}
%%%%%%%%%%%%%%%%%%%%%%%%%%%%%%%%%%%%%%%%%%%%%%%%%%%%%%%%%%%%%%

For a quantitative discussion of the physical behaviour of the 
HBT radius parameters, in particular of their ${\bf K}$-dependence, 
we use a simple model for the emission function of a finite expanding 
thermalized source \cite{CNH95}:
 \begin{equation}
 \label{3.15}
   S(x,K) = {M_\perp \cosh(\eta-Y) \over
            (2\pi)^3 \sqrt{2\pi(\Delta \tau)^2}}
            \exp \left[- {K \cdot u(x) \over T}
                       - {(\tau-\tau_0)^2 \over 2(\Delta \tau)^2}
                       - {r^2 \over 2 R^2} 
                       - {{\eta- \eta_0}^2 \over 2 (\Delta \eta)^2}
           \right] .
 \end{equation}
Here $r = \sqrt{x^2+y^2}$, the spacetime rapidity $\eta = {1 \over 2} 
\ln[(t+z)/(t-z)]$ and the longitudinal proper time $\tau= \sqrt{t^2-
z^2}$ parametrize the spacetime coordinates $x^\mu$, with measure 
$d^4x = \tau\, d\tau\, d\eta\, r\, dr\, d\phi$.  $Y = {1\over 2} 
\ln[(1+\beta_l)/(1-\beta_l)]$ and $M_\perp = \sqrt{m^2 + K_\perp^2}$ 
parametrise the longitudinal and transverse components of the pair 
momentum ${\bf K}$. 

\vspace*{9cm}
%\special{psfile=kaon.ps hoffset=20 voffset=60 hscale=40 vscale=45 angle=0}
\includegraphics{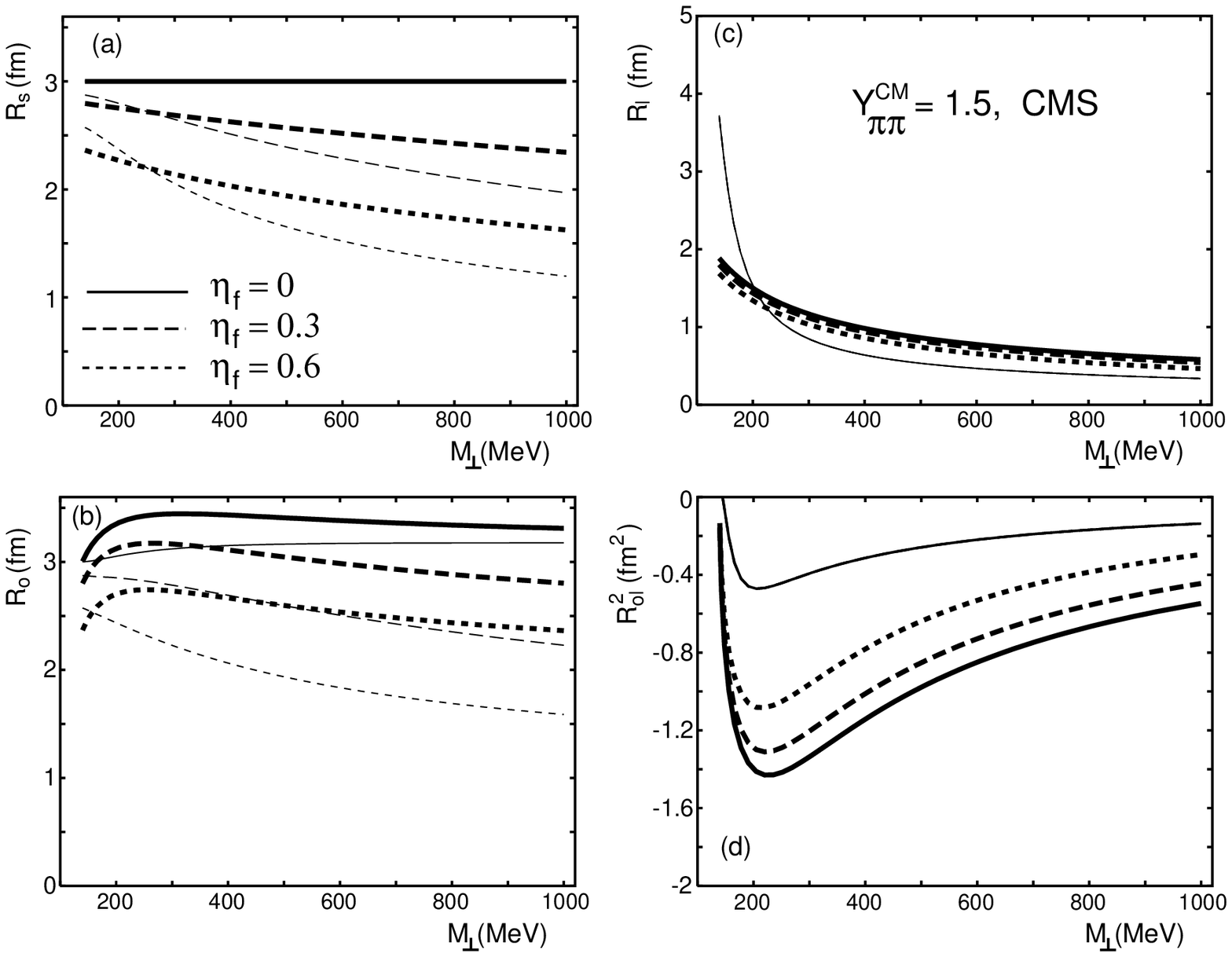}
%\vskip -5cm
\begin{center}
\begin{minipage}[t]{13cm}
\noindent \bf Fig.1.  \rm
The standard Cartesian parameters $R_s$ (a), $R_o$ (b), $R_l$ (c),
and $R_{ol}^2$ (d) in the CMS for pion pairs with c.m. rapidity $Y=1.5$,
as functions of $M_\perp$ for 3 different values for the transverse 
flow $\eta_f$. The thick lines are exact numerical results from 
Eqs.~(\protect\ref{16a}-\protect\ref{16d}), the thin lines are obtained 
from the analytical approximations given in Ref.~\protect\cite{CL95}. 
(Figure taken from Ref.~\protect\cite{TWH96}.)
\end{minipage}
\end{center}
%\vskip 4truemm

\noindent $T$ is the freeze-out temperature, $R$ is the 
transverse geometric (Gaussian) radius of the source, $\tau_0$ its 
average freeze-out proper time, $\Delta \tau$ the mean proper time 
duration of particle emission, and $\Delta \eta$ parametrises 
the finite longitudinal extension of the source. The 
expansion flow velocity $u^\mu(x)$ is parametrised as 
 \begin{equation}
 \label{26}
   u^\mu(x){=}\left( \cosh \eta_l \cosh \eta_t(r), 
                     \sinh \eta_t(r)\, {\bf e}_r,  
                     \sinh \eta_l \cosh \eta_t(r) \right) ,
  \  \eta_l{=}\eta ,
  \  \eta_t(r){=}\eta_f (r/R)  ,
 \end{equation}
with a boost-invariant longitudinal flow rapidity and a linear 
transverse flow rapidity profile. $\eta_f$ scales the strength of 
the transverse flow. The scalar product in the exponent of the 
Boltzmann factor can then be written as
  \begin{equation}
    \label{2.5}
    K\cdot u(x) = M_\perp \cosh(\eta - Y) \cosh\eta_t(r) 
                - K_\perp {x\over r} \sinh\eta_t(r) \, .
  \end{equation}
Please note that for non-zero transverse momentum $K_\perp$, a finite 
transverse flow breaks the azimuthal symmetry of the emission function 
via the second term in (\ref{2.5}). For $\eta_f=0$ the source has no 
explicit $K_\perp$-dependence, and $M_\perp$ is the only relevant scale. 
As will be discussed in Sec.~\ref{sec5c} this gives rise to perfect 
$M_\perp$-scaling of the YKP radius parameters in the absence of 
transverse flow, which is again broken for non-zero transverse flow 
\cite{HTWW96a}. 

For the numerical calculations below we have selected one fixed set of 
source parameters: $R=3$ fm, $\tau_0 = 3$ fm/$c$, 
$\Delta \tau = 1$ fm/$c$, $\Delta \eta = 1.2$, $T=140$ MeV.

\vspace*{9cm}
%\special{psfile=kaon.ps hoffset=20 voffset=60 hscale=40 vscale=45 angle=0}
\includegraphics{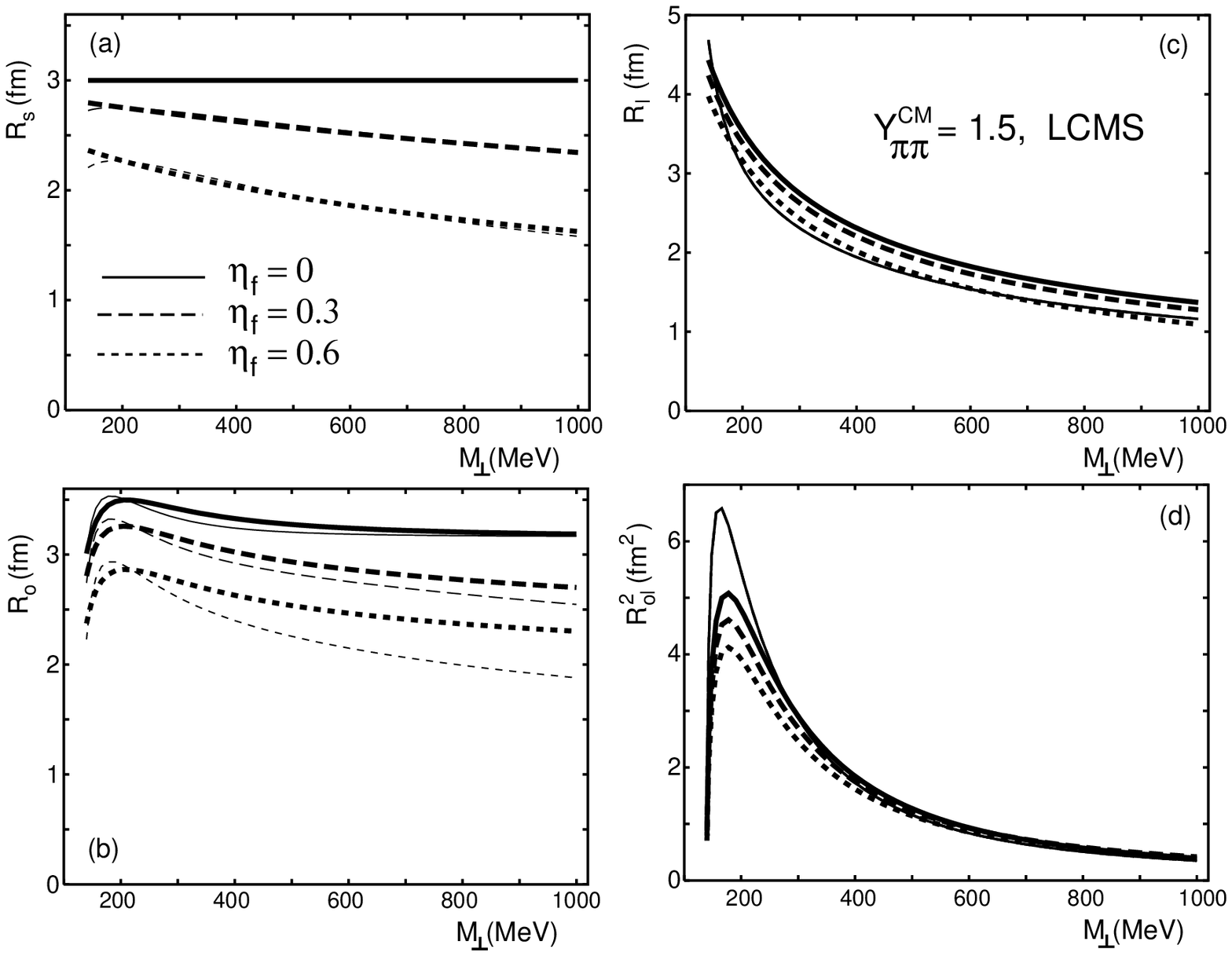}
%\vskip -5cm
\begin{center}
\begin{minipage}[t]{13cm}
\noindent \bf Fig.2.  \rm
Same as Fig.1, but now evaluated in the LCMS. Please note the change of sign 
and magnitude of the cross-term. 
(Figure taken from Ref.~\protect\cite{TWH96}.)
\end{minipage}
\end{center}
%\vskip 4truemm

%%%%%%%%%%%%%%%%%%%%%%%%%%%%%%%%%%%%%%%%%%%%%%%%%%%%%%%%%%%%%%
\section{MOMENTUM DEPENDENCE OF HBT PARAMETERS}
\label{sec5}
%%%%%%%%%%%%%%%%%%%%%%%%%%%%%%%%%%%%%%%%%%%%%%%%%%%%%%%%%%%%%%

%%%%%%%%%%%%%%%%%%%%%%%%%%%%%%%%%%%%%%%%%%%%%%%%%%%%%%%%%%%%%%
\subsection{Standard Cartesian fit}
\label{sec5a}
%%%%%%%%%%%%%%%%%%%%%%%%%%%%%%%%%%%%%%%%%%%%%%%%%%%%%%%%%%%%%%
 
In Fig.~1 I show the HBT radius parameters from the standard Cartesian fit
(\ref{15}) for pion pairs with c.m. rapidity $Y=1.5$ where the fit of the
correlator is done in the CMS \cite{TWH96}. The different thick curves 
correspond to different strengths $\eta_f$ of the transverse flow. Without 
transverse flow $R_s$ is $M_\perp$-independent because the source 
(\ref{3.15}) has no transverse temperature 
gradients. As transverse flow increases, $R_s$ develops an increasing 
dependence on $M_\perp$. It can be approximated by an inverse power law,
with the power increasing monotonously with $\eta_f$ \cite{WSH96,WHTW96}.
$R_l$ features a very strong $M_\perp$-dependence even without transverse 
flow, due to the strong longitudinal expansion of the source. It can also
be described by an inverse power law, with a larger power $\simeq 0.55$,
in rough agreement with the approximate $\sqrt{T/M_\perp}$-scaling law
suggested in \cite{MS88} (see, however, \cite{WSH96,HB95} for a more 
quantitative discussion). The increase of $R_o$ at small $M_\perp$ is due 
to the contribution (\ref{17}) from the effective lifetime.
As seen in Fig.~4 below, in the YK frame (source rest frame) the latter 
is of order 2.5 fm/$c$ at small $M_\perp$; Fig.~1b shows that its effect 
on $R_o$ compared to $R_s$ in the CMS is much smaller (and thus more 
difficult to measure). Fig.~1d shows that the cross-term is small in the CMS
but non-zero. It vanishes at $K_\perp=0$ by symmetry and also becomes 
small again at large $K_\perp$. 

The thin lines in Fig.~1 show for 
comparison approximate results for the HBT radii calculated from the 
approximate analytical results given in Ref.~\cite{CL95} which were 
derived by evaluating Eqs.~(\ref{16a}-\ref{16d}) by saddle point 
integration. It is clear that this method fails here (see Ref.~\cite{WSH96}
for a quantitative discussion of this approximation), and that the 
analytical expressions should not be used for a quantitative analysis of 
HBT data. 

\vspace*{6cm}
%\special{psfile=kaon.ps hoffset=20 voffset=60 hscale=40 vscale=45 angle=0}
\includegraphics{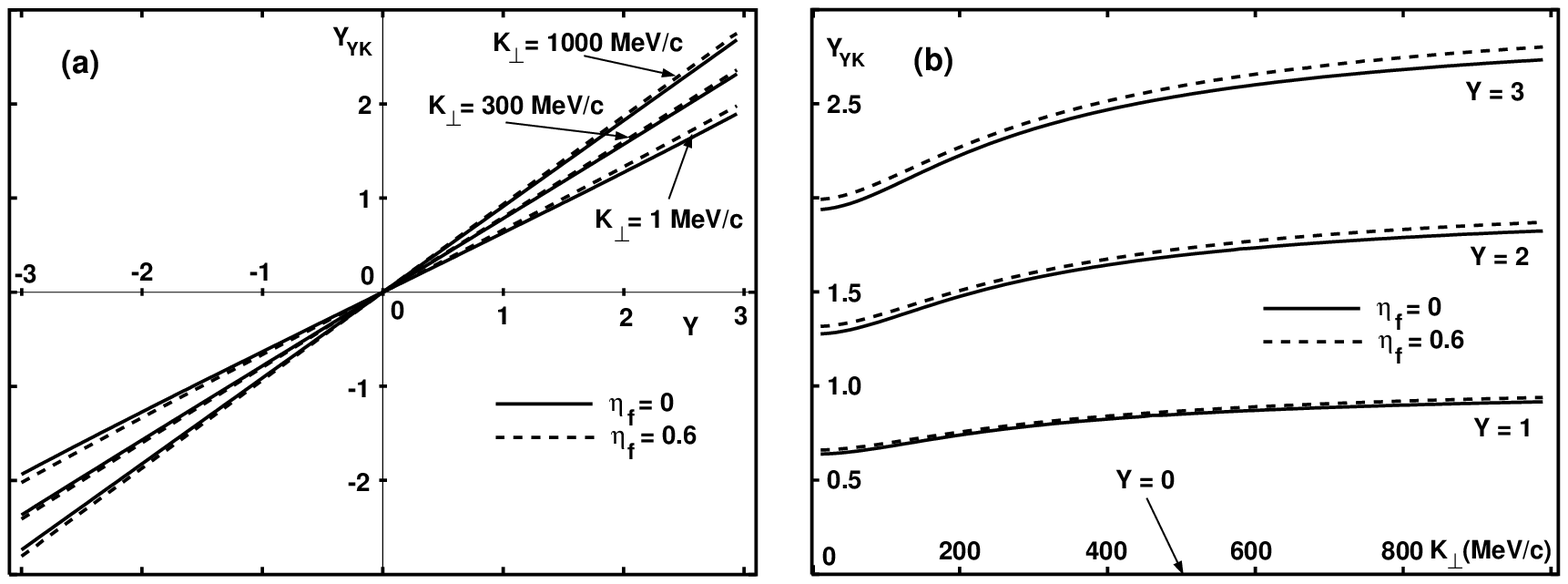}
%\vskip -5cm
\begin{center}
\begin{minipage}[t]{13cm}
\noindent \bf Fig.3.  \rm
(a) The Yano-Koonin rapidity for pion pairs, as a function of the pair 
c.m. rapidity $Y$, for various values of $K_\perp$ and two values for 
the transverse flow $\eta_f$. (b) The same, but plotted against $K_\perp$ 
for various values of $Y$ and $\eta_f$.
(Figure taken from Ref.~\protect\cite{HTWW96}.)
\end{minipage}
\end{center}
%\vskip 4truemm

Fig.~2 shows the same situation as Fig.~1, but now all HBT radii are 
evaluated in the LCMS (longitudinally comoving system \cite{CP91}) which 
moves with the pair rapidity $Y=1.5$ relative to the CMS. A comparison with 
Fig.~1 shows the strong reference frame dependence of the standard HBT 
radii. In particular, the cross-term changes sign and is now much larger.
The analytical approximations from Ref.~\cite{CL95} work much better in 
the LCMS \cite{CL95}, but for $R_o$ and $R_{ol}^2$ they are still not 
accurate enough (in particular in view of the delicate nature of the 
lifetime effects on $R_o$).

%%%%%%%%%%%%%%%%%%%%%%%%%%%%%%%%%%%%%%%%%%%%%%%%%%%%%%%%%%%%%%
\subsection{The Yano-Koonin velocity}
\label{sec5b}
%%%%%%%%%%%%%%%%%%%%%%%%%%%%%%%%%%%%%%%%%%%%%%%%%%%%%%%%%%%%%%

Fig.~3 shows (for pion pairs) the dependence of the YK velocity on the 
pair momentum ${\bf K}$. In Fig.~3a we show the YK rapidity $Y_{_{\rm YK}} =
\frac 12 \ln[(1+v)/(1-v)]$ as a function of the pair rapidity $Y$
(both relative to the CMS) for different values of $K_\perp$,
in Fig.~3b the same quantity as a function of $K_\perp$ for different $Y$.
Solid lines are without transverse flow, dashed lines are for $\eta_f=0.6$.
For large $K_\perp$ pairs, the YK rest frame approaches the LCMS (which 
moves with the pair rapidity $Y$); in this limit all pairs are thus 
emitted from a small region in the source which moves with the same 
longitudinal velocity as the pair. For small $K_\perp$ the YK frame 
is considerably slower than the LCMS; this is due to the thermal 
smearing of the particle velocities in our source around the local 
fluid velocity $u^\mu(x)$ \cite{WHTW96}. The linear relationship between 
the rapidity $Y_{_{\rm YK}}$ of the Yano-Koonin frame and the pion pair 
rapidity $Y$ is a direct reflection of the boost-invariant longitudinal 
expansion flow \cite{HTWW96}. For a non-expanding source $Y_{_{\rm YK}}$ 
would be independent of $Y$. Additional transverse flow is seen to have 
nearly no effect. The dependence of the YK velocity on the pair rapidity thus
measures directly the longitudinal expansion of the source and cleanly 
separates it from its transverse dynamics. A detailed discussion of these 
features is given in Ref.~\cite{WHTW96}.

%%%%%%%%%%%%%%%%%%%%%%%%%%%%%%%%%%%%%%%%%%%%%%%%%%%%%%%%%%%%%%
\subsection{$M_\perp$-scaling of YKP radii and transverse flow}
\label{sec5c}
%%%%%%%%%%%%%%%%%%%%%%%%%%%%%%%%%%%%%%%%%%%%%%%%%%%%%%%%%%%%%%

In the absence of transverse flow, a thermal source like (\ref{3.15}) 
depends on the particle rest mass and on the transverse momentum 
$K_\perp$ only through the combination $M_\perp^2 = m^2 +K_\perp^2$ (see 
Eq.~(\ref{2.5})). Furthermore, the source
is then azimuthally and $x\to -x$ reflection symmetric. Hence $\langle
\tilde x \tilde t \rangle$, $\langle \tilde x \tilde z \rangle$, and 
$\langle \tilde x^2 - \tilde y^2\rangle$ all vanish and the approximations 
in Eqs.~(\ref{20b},\ref{20c}) become exact. As a result, all three YKP
radii (\ref{20a})-(\ref{20c}) are only functions of $M_\perp$, too 
(as well as of Y, of course), i.e. they do not depend explicitly on the 
particle rest mass. 

\vspace*{12cm}
%\special{psfile=kaon.ps hoffset=20 voffset=60 hscale=40 vscale=45 angle=0}
\includegraphics{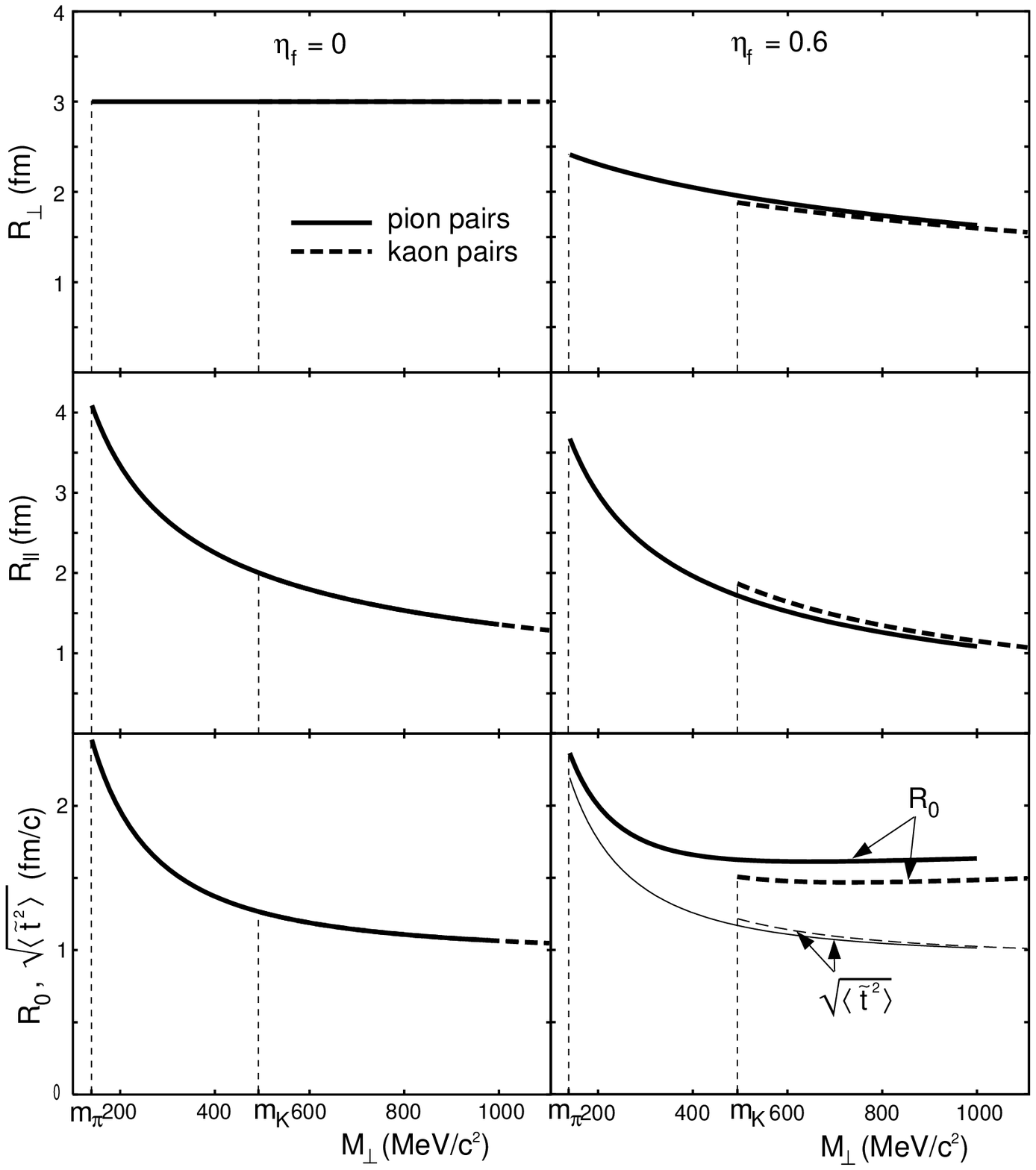}
%\vskip -5cm
\begin{center}
\begin{minipage}[t]{13cm}
\noindent \bf Fig.4.  \rm
The YKP radii $R_\perp$, $R_\parallel$, and $R_0$ (from top to bottom)
for vanishing transverse flow (left column) and for $\eta_f=0.6$ (right
column), as functions of $M_\perp$ for pairs at $Y_{\rm cm}=0$. 
Solid (dashed) lines are for pions (kaons). The breaking of the
$M_\perp$-scaling by transverse flow is obvious in the right column.
Also, as shown in the lower right panel, for nonzero transverse flow 
$R_0$ does not agree exactly with the effective source lifetime 
$\protect\sqrt{\langle \tilde t^2\rangle}$. 
(Figure taken from Ref.~\protect\cite{WHTW96}.)
\end{minipage}
\end{center}
%\vskip 4truemm

This is seen in the left column of Fig.~4 where the three
YKP radii are plotted for $Y_{\rm cm}=0$ pion and kaon pairs as functions 
of $M_\perp$; they agree perfectly. 
The transverse radius here shows no $M_\perp$-dependence due to the 
absence of transverse temperature gradients, but even with temperature 
gradients it would only depend on $M_\perp$.
(Of course, this discussion neglects resonance decays which will
be studied in Sec.~\ref{sec6}.) Note that $M_\perp$-scaling 
in the absence of transverse flow applies only to the YKP radius parameters:
since the expressions (\ref{16b})-(\ref{16d}) involve nonvanishing 
variances with $\beta_\perp$- or $\beta_l$-prefactors (which depend 
explicitly on the rest mass), the HBT radii from the standard Cartesian 
fit do not exhibit $M_\perp$-scaling.

For non-zero transverse flow $\eta_f\ne 0$ this $M_\perp$-scaling is 
broken by two effects: first, the second term in (\ref{2.5}) destroys
the $M_\perp$-scaling of the emission function itself, and second
the $\bbox{\beta}$-dependent correction terms in (\ref{20b},\ref{20c}) 
are now non-zero because the same term also breaks, for $K_\perp\ne 0$,
the $x \to -x$ and $x \to y$ symmetries. The magnitude of the associated
scale breaking due to the pion-kaon mass difference is seen in the right 
column of Fig.~4 for $\eta_f=0.6$. The effects are small and require very
accurate experiments for their detection. However, the sign of the effect
is opposite for $R_\parallel$ and for $R_\perp,\, R_0$ which may help 
to distinguish flow-induced effects from resonance decay contributions. 

Since for $Y_{\rm cm}=0$ the YK and CMS frames coincide, $\beta_l=0$ in 
the YK frame and the approximation in (\ref{20b}) remains exact even 
for non-zero transverse flow. The same is not true for the approximation 
in (\ref{20c}), and therefore we show in the lower right panel of Fig.~4 
also the effective source lifetime $\sqrt{\langle \tilde t^2 \rangle}$
for comparison. The apparently rather large discrepancies between
the YKP parameter $R_0$ and the effective source lifetime is due to
a rather extreme choice of parameters: a large flow transverse flow and
a small intrinsic source lifetime of $\Delta\tau = 1$ fm/$c$ in (\ref{3.15}).
Since $\sqrt{\langle \tilde t^2 \rangle}$ approaches $\Delta\tau$ in the limit
of large $M_\perp$ while the dominant \cite{WHTW96} correction term 
$\langle \tilde x^2 - \tilde y^2 \rangle$ does not depend on $\Delta\tau$,
the YKP parameter $R_0$ will track the effective source lifetime more 
accurately for larger values of $\Delta\tau$ (and for smaller values 
of $\eta_f$).

Why do $\sqrt{\langle \tilde t^2 \rangle}$ and $R_0$ increase at small 
$M_\perp$? Due to the rapid longitudinal expansion, the longitudinal region
of homogeneity $R_\parallel$ is a decreasing function $M_\perp$.
Since for different pair momenta $R_0$ measures the source lifetime
in different YK reference frames, the freeze-out ``hypersurface'' will
in general appear to have different shapes for pairs with different momenta.
Only in our model, where freeze-out occurs at fixed proper time $\tau_0$
(up to a Gaussian smearing with width $\Delta\tau$), is it frame-independent.
It is thus generally unavoidable (and here, of course, true in any frame)
that freeze-out at different points $z$ in the source will occur at different
times $t$ in the YK frame. Since a $z$-region of size $R_\parallel$ 
contributes to the correlation function, $R_\parallel$ determines how large 
a domain of this freeze-out surface (and thus how large an interval of
freeze-out times in the YK frame) is sampled by the correlator. This
interval of freeze-out times combines with the intrinsic Gaussian width 
$\Delta\tau$ to yield the total effective duration of particle emission.
It will be largest at small pair momenta where the homogeneity region 
$R_\parallel$ is biggest, and will reduce to just the variance of the
Gaussian proper time distribution at large pair momenta where the
longitudinal (and transverse) homogeneity regions shrink to zero. The rise
of $\Delta t({\bf K})$ at small ${\bf K}$ is thus generic.

%%%%%%%%%%%%%%%%%%%%%%%%%%%%%%%%%%%%%%%%%%%%%%%%%%%%%%%%%%%%%%
\section{RESONANCE DECAYS}
\label{sec6}
%%%%%%%%%%%%%%%%%%%%%%%%%%%%%%%%%%%%%%%%%%%%%%%%%%%%%%%%%%%%%%

The proportionality of the $M_\perp$-dependence of $R_\perp$ to the 
transverse flow $\eta_f$ and the particular pattern of $M_\perp$ 
scale-breaking by the latter open an avenue for the quantitative 
extraction of transverse flow from HBT data \cite{HTWW96a}. This 
requires, however, that the $M_\perp$-dependence of $R_\perp$ is not affected
by resonance decays. Since they contribute more to pions than 
to kaons they may also affect the $M_\perp$-scaling arguments. 
The work by the Marburg group \cite{Marb} on resonance decay effects on HBT
in the context of hydrodynamical simulations indicates, within the
standard Cartesian framework and without accounting for the cross-term,
a possible additional $M_\perp$-dependence of the transverse radius.
However, a systematic analysis of resonance contributions to HBT as a 
function of various characteristic source parameters is only now
becoming available \cite{WH96a}.

\vspace*{10cm}
%\special{psfile=kaon.ps hoffset=20 voffset=60 hscale=40 vscale=45 angle=0}
\includegraphics{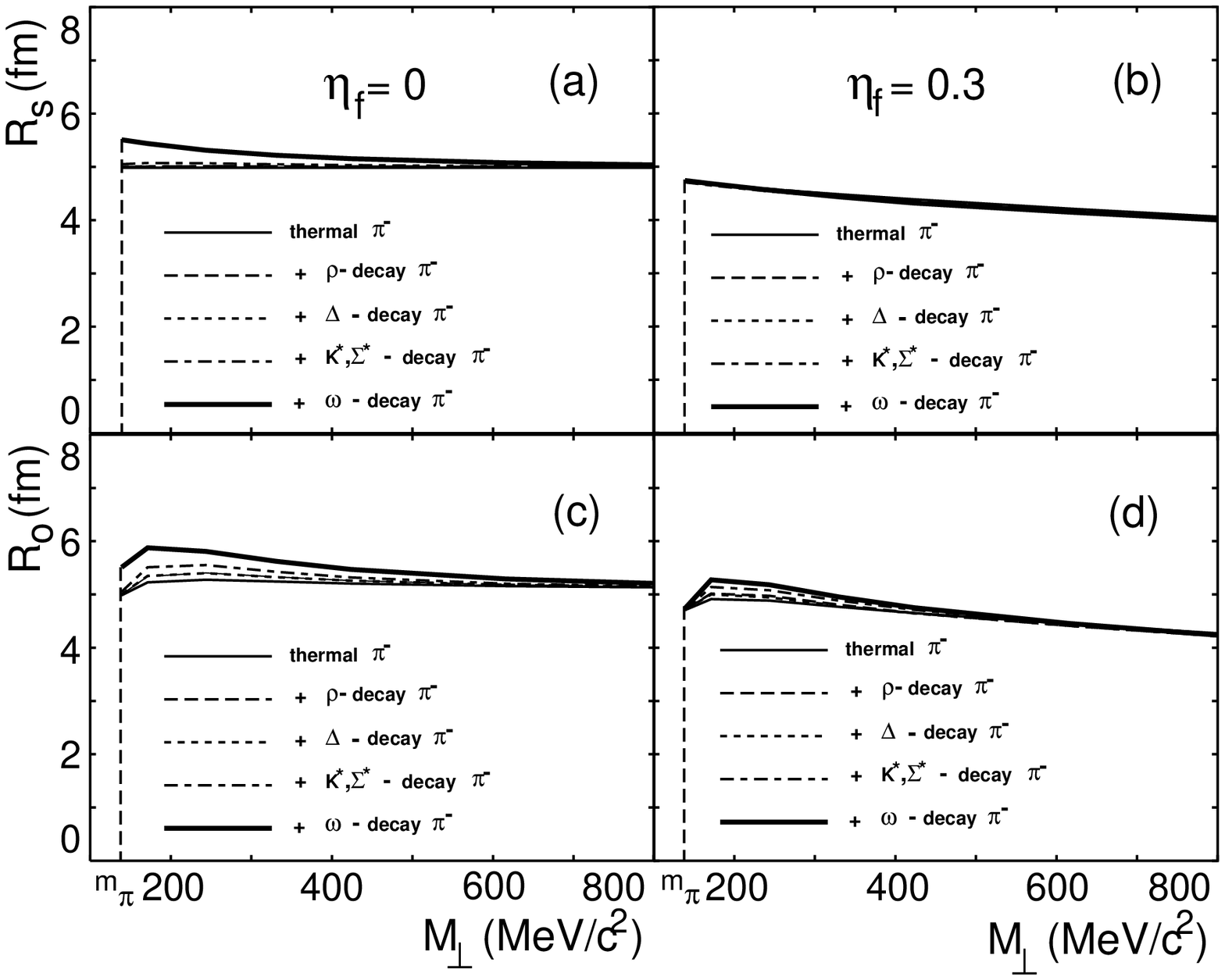}
%\vskip -5cm
\begin{center}
\begin{minipage}[t]{13cm}
\noindent \bf Fig.5.  \rm
The influence of resonance decays on the $M_\perp$-dependence of
$R_s$ (a,b) and $R_o$ (c,d) for $Y_{\rm cm}=0$ pion pairs. a,c: no 
transverse flow; b,d: transverse flow rapidity $\eta_f=0.3$.
The Gaussian transverse radius is here $R=5$ fm, and $T=150$ MeV.
(Figure taken from Ref.~\protect\cite{WH96a}.)
\end{minipage}
\end{center}
%\vskip 4truemm

In Fig.~5 I show some results from Ref.~\cite{WH96a} for the same 
emission function (\ref{3.15}). The only change for resonances
is an additional spin degeneracy factor and the different rest mass. 
The complete spectrum of relevant resonances is included, and in the
decays the 2- and 3-body decay kinematics is fully taken into 
account. The HBT radii are extracted from a Gaussian fit to the 
numerically calculated correlation function. A detailed technical 
discussion is given in Ref.~\cite{WH96a}.

Fig.~5 shows that the effects of the short-lived resonances with lifetimes
of order 1 fm/$c$ on $R_s$ are essentially negligible, both at vanishing 
and at nonzero transverse flow. Only the $\omega$ with its intermediate 
lifetime of 20 fm/$c$ affects $R_s$, but only for vanishing transverse flow. 
There it induces a weak $M_\perp$-dependence at small $M_\perp$ even in 
the absence of transverse flow; at $M_\perp>500$ MeV the contribution
of the $\omega$ dies out, and $R_s$ again becomes $M_\perp$-independent 
(which would not be the case if it were affected by flow). At $\eta_f=0.3$ 
and 0.6 \cite{WH96a} not even the $\omega$ generates any additional 
$M_\perp$-dependence! --
$R_o$ shows some effects from the additional lifetime of the resonances,
in particular from the long-lived $\omega$. Resonances with much longer 
lifetimes than the $\omega$ (in particular all weak decays) have 
no effect on the radii, because their contribution to the correlator is 
only at very small values of $q$ which cannot be resolved experimentally. 
They lead to a reduced ``incoherence parameter'' $\lambda$ 
\cite{Marb,CLZ94}. Since for increasing $M_\perp$ the resonance 
contributions decrease, the $\lambda$-parameter increases with $M_\perp$, 
approaching 1 as $M_\perp\to\infty$ \cite{Marb,CLZ94}. A detailed study 
will follow \cite{WH96a}.

The weak effect of resonances on $R_s=R_\perp$ seems surprising: due 
to their non-zero lifetime they should be able to propagate outside the 
original source before decay and form a pion ``halo'' \cite{Marb,CLZ94}.
This effect is, however, much weaker than naively expected: most
of the resonances are not very fast, and the halo thickness is thus only
a fraction of the resonance lifetime. At finite transverse flow an
additional effect comes into play: it turns out that then the effective 
size of the emission function for directly emitted resonances is 
{\em smaller} than that for direct pions \cite{WH96a}! At $\eta_f{=}0.3$ 
and 0.6 this even slightly overcompensates the halo effect, and altogether 
the resonances change neither the size nor the $M_\perp$-dependence 
of $R_s$. 

%%%%%%%%%%%%%%%%%%%%%%%%%%%%%%%%%%%%%%%%%%%%%%%%%%%%%%%%%%%%%%
\section{CONCLUSIONS}
\label{sec7}
%%%%%%%%%%%%%%%%%%%%%%%%%%%%%%%%%%%%%%%%%%%%%%%%%%%%%%%%%%%%%%

The model-independent expressions of Secs.~\ref{sec2} and \ref{sec4}
for the HBT width
parameters in terms of second order variances of the emission function
provide the basis of a detailed physical interpretation of the measured 
HBT radii. They show that the HBT radius parameters do not necessarily
measure the full geometric extension of the source, but regions of 
homogeneity in the effective emission function for particles with 
certain fixed momenta. For expanding systems these are usually smaller
than the naive geometric source size and decreasing functions of the 
pair momentum. For systems with finite lifetime the HBT parameters
usually mix the spatial and temporal structure of the source, and their 
unfolding requires model studies. 

With the new YKP parametrization we have found a method which, for systems 
with dominant longitudinal expansion, cleanly factorises the longitudinal 
and transverse spatial from the temporal homogeneity length. The effective 
source lifetime is directly fitted by the parameter $R_0$; it is generically
a function of the pair momentum and largest for pairs which are slow 
in the CMS. Another fit parameter, the YK velocity, measures directly 
the longitudinal velocity of the emitting fluid element, and its 
dependence on the pair rapidity allows for a direct determination
of the longitudinal expansion of the source. Without transverse expansion, 
the YKP radius parameters show exact $M_\perp$-scaling. The breaking of this
scaling and the $M_\perp$-dependence of the transverse radius parameter
$R_\perp$ allow for a determination of the transverse expansion velocity
of the source. Resonance decays were shown to mostly affect the lifetime
parameter and leave the $M_\perp$-dependence of $R_\perp$ nearly unchanged.
They thus do not endanger the extraction of the transverse flow via
HBT.

\vskip 0.2cm

With this new and detailed understanding of the method, I believe 
that HBT interferometry has a begun a new and vigorous life as a 
powerful tool for reconstructing the geometric and dynamic space-time 
characteristics of the collision zone from the measured momentum spectra. 

\vskip 0.4cm

\noindent {\bf Acknowledgements:} 
I thank my collaborators on this project, S. Chapman, J.R. Nix, B. 
Tom\'a\v sik, U.A. Wiedemann, and Wu Yuanfang, who each contributed 
valuable pieces to the puzzle. Without their help until the very last 
minutes before the conference this review would have been impossible. 
I would also like to acknowledge fruitful discussions with H. 
Appelsh\"auser, T. Cs\"org\H o, D. Ferenc, M. Ga\'zdzicki, and P. 
Seyboth. This work was supported by grants from BMBF, DFG, and GSI.

%\newpage

%%%%%%%%%%%%%%%%%%%%%%%%%%%%%%%%%%%%%%%%%%%%%%%%%%%%%%%%%%%%%%%%%%%%
% References:
%%%%%%%%%%%%%%%%%%%%%%%%%%%%%%%%%%%%%%%%%%%%%%%%%%%%%%%%%%%%%%%%%%%%

\end{document}